\documentclass[prl, amsmath,twocolumn, amssymb,superscriptaddress]{revtex4-1}

\usepackage{amsmath}
\usepackage{amssymb}
\usepackage{amsthm}
\usepackage[pdftex]{color}
\usepackage{graphicx}
\usepackage{dcolumn} 
\usepackage{bm} 
\usepackage{epic}
\usepackage{longtable}
\usepackage{ulem}   
\newcommand{\pdag}{{\phantom{\dag}}}

\def \be{\begin{equation}}
\def \ee{\end{equation}}
\def \ba{\begin{array}}
\def \ea{\end{array}}
\def \bea{\begin{eqnarray}}
\def \eea{\end{eqnarray}}

\def \e{{\epsilon}}

\def \a{{\alpha}}

\def \b{{\beta}}
\def \g{{\gamma}}
\def \D{{\Delta}}

\def \e{{\epsilon}}

\def \yd{^\dagger}
\def \av#1{{\langle#1\rangle}}

\begin{document}
\title{
 Dynamics of a Many-Body-Localized System Coupled to a Bath 
      }
\author{Mark H Fischer}
\author{Mykola Maksymenko}
\author{Ehud Altman}
\affiliation{Department of Condensed Matter Physics, Weizmann Institute of Science, Rehovot 761001, Israel}


\begin{abstract}
  Coupling a many-body-localized system to a dissipative bath necessarily leads to delocalization.
  Here, we investigate the nature of the ensuing relaxation dynamics and the information it holds on the many-body-localized state. 
  We formulate the relevant Lindblad equation in terms of the local integrals of motion of the underlying localized Hamiltonian. 
  This allows to map the quantum evolution deep in the localized state to tractable classical rate equations. 
  We consider two different types of dissipation relevant to systems of ultra-cold atoms: dephasing due to inelastic scattering on the lattice lasers and particle loss. 
Our approach allows us to characterize their different effects in the limiting cases of weak and strong interactions.
\end{abstract}

\maketitle

\emph{Introduction --}
A closed quantum system can fail to reach thermal equilibrium when evolving under its own intrinsic dynamics if it is (many-body) localized by strong quenched disorder~\cite{basko:2006}. Direct evidence for such breaking of ergodicity in an interacting many-body system 
has recently been observed in a system of ultra-cold atomic fermions~\cite{schreiber:2015, bordia:2015tmp}, as well as a chain of trapped ions~\cite{smith:2015tmp}. However, even these highly controlled implementations of many-body-localized (MBL) systems are affected by at least small coupling to the environment, for example through inelastic scattering on the lasers, that ultimately destroys the localization. Therefore, in order to study the phenomenon of many-body localization in a realistic setting it is important to address the effect of dissipation on such systems and understand the interplay with intrinsic localization properties of the many-body system.

Here, we develop a general framework to investigate the dynamics of a MBL system weakly coupled to a dissipative Markovian bath. Our approach is based on the local integrals of motion of the closed localized system. Using our general framework we discuss how observables relax in the presence of different types of dissipative baths. We note that earlier studies discussed the effects of a coupling to a thermal bath on the spectral properties of a MBL system~\cite{nandkishore:2014b, Johri:2015}.
A particular example of dynamical relaxation of observables in a MBL system coupled to a bath has also been addressed recently in a numerical study~\cite{levi:2015tmp}, where a slow stretched exponential decay of these observables was found in a wide regime. We explain the origin of this behavior and address how the relaxation properties depend on the interactions and on the type of bath affecting the system. 

\emph{General scheme --}
For concreteness, we exemplify our general approach using a one-dimensional model of spinless lattice fermions with random on-site energies and nearest-neighbor interactions, 
\begin{equation}
  H = -J\sum_i (c^{\dag}_i c^{\phantom{\dag}}_{i+1} + h.c.) + \sum_i V_i n_i + U\sum_{i} n_in_{i+1}.
  \label{eq:HIntAnderson}
\end{equation}
Here, $c^\dagger_i$ creates a fermion on site $i$, $n_i = c_i^{\dag}c^\pdag_i$, and the on-site energies  $V_i \in [-h, h]$ are independent random variables. As a model of a closed system, this Hamiltonian has been extensively studied and (for $U=2J$) is expected to be localized for $h\gtrsim 7.2$~\cite{pal:2010, luitz:2015, serbyn:2015}.

The time evolution of the density matrix of the system coupled to a Markovian bath is given by the Lindblad equation 
\begin{equation}
  \dot{\rho} = \mathcal{L}[\rho] = -i[H, \rho] + \gamma \sum_{i}\Big(L^\pdag_{i}\rho L_{i}^\dag - \frac12\{L_{i}^\dag L_{i}^\pdag, \rho\}\Big),
  \label{eq:lindblad}
\end{equation}
where the first term governs the unitary time evolution while the second term describes the coupling of the system to the environment after tracing out the bath degrees of freedom. $L_i$ are called jump operators, the system operators directly coupled to the bath. Within a stochastic quantum-trajectory picture the unitary evolution of an ensemble of pure states is interrupted at random times by non-unitary quantum measurements (or jumps) generated by the operators $L_i$~\cite{dalibard:1992, carmichael:1993}.

A full solution of the Lindblad dynamics \eqref{eq:lindblad} is a daunting quantum problem. However, we are able to greatly simplify it in the (fully) MBL phase, where the Hamiltonian can be represented in terms of a complete set of quasi-local integrals of motion (LIOMs)~\cite{serbyn:2013, huse:2014, ros:2015, chandran:2015}. 
The corresponding operators $\tilde{n}_i$ are in one-to-one correspondence and have a finite overlap with the truly local physical operators $n_i$ appearing in the original Hamiltonian. 

We utilize the fact that the unitary evolution deep in the MBL state is trivial in the LIOM basis. For physical initial conditions (i.e. local in the physical basis $n_i$) the off-diagonal elements of the density matrix in this basis dephase rapidly. Hence, after a microscopic time scale (of order $1/h$) the density matrix becomes effectively diagonal. Within the quantum-trajectory picture, the jump operators $L_i$ act after random time intervals set by $1/\gamma$ and project to a new state $\rho$ with off-diagonal elements in the LIOM basis. However, for sufficiently weak coupling ($\gamma/ h\ll 1$), the density matrix has enough time to dephase between successive quantum jumps as local operators $L_i$ do not create resonances deep in the localized phase. We can thus view the density matrix as a classical (joint) probability distribution $\rho(\{\tilde{n}_i\})$ at all times. 

The time evolution of the probability  distribution is given by a classical master equation, where the quantum jumps generate the transition rates between classical configurations. Hence, the change of the probability distribution of some configuration $a$ is just given by the transitions in and out of it,
\begin{equation}
  {d\rho(a)\over dt}=\sum_b \left[W(b,a) \rho(b)-W(a,b)\rho(a)\right],
  \label{eq:classicalrate}
\end{equation}
where $a$ and $b$ represent classical configurations $\{\tilde{n}_1,\ldots,\tilde{n}_L\}$ and $W(b,a)=\gamma \sum_i|\langle a|L_i| b\rangle|^2$ is the transition rate from $b$ to $a$. These rates are obtained by writing the jump operators in terms of the LIOMs.
Note that for the problem of spinless Fermions the classical master equation then describes a symmetric exclusion process~\cite{liggett:2004}. For most of the conditions we consider, namely high temperature and/or weak interactions, a simple mean-field solution of the classical problem is justified and, as we show, gives an excellent approximation of the full quantum dynamics. 

Motivated by the recent experiment of Ref.~\cite{schreiber:2015} we initialize the system in a state with a period-two density wave and track the relaxation of the density imbalance $\mathcal{I}(t) = \langle n_e - n_o\rangle/\langle n_e + n_o \rangle$ with $n_{e}$ ($n_o$) the occupation of even (odd) sites. Without coupling to a bath, $\mathcal{I}(t)$ approaches a non-vanishing value in the MBL phase, while it decays completely on a short timescale $\sim 1/J$ in the ergodic phase.
We then consider two types of dissipative mechanisms that would ensue due to inelastic scattering of the atoms on the laser fields in the above mentioned experiment. 
The first mechanism is recoil and dephasing of the atom wavefunction implemented by a local measurement of the on-site density, $L_i = n_i$~\cite{Pichler2010}.
The second one is particle loss, which is implemented by the jump operators $L_i = c_i$. 

\emph{Non-interacting system --}
Having described the general approach we now demonstrate it for the case of a non-interacting system [$U=0$ in Eq.~\eqref{eq:HIntAnderson}]. The LIOMs in this case are simply the occupation numbers of the localized single-particle states $\tilde{n}_\alpha=\tilde{c}^\dag_\alpha\tilde{c}^\pdag_\alpha$, where $\tilde{c}^\dag_\alpha = \sum_i \phi^*_{\alpha i}c^\dag_i$ 
 creates a fermion in the localized state $\a$. The indices $\alpha$ are in one-to-one correspondences with the site indices $i$, so that $\phi_{\alpha i}$ are the single-particle wave functions localized near sites $\a$.

We follow the procedure outlined above considering first the effect of a dephasing bath, i.e., $L_i=n_i$. These jump operators are written in terms of the ``quasi-particle'' operators $\tilde{c}_\a$ as $L_i=\sum_{\mu\nu}\phi^{\phantom{*}}_{\mu i}\phi^{*}_{\nu i} \tilde{c}\yd_\mu\tilde{c}^\pdag_\nu$. This representation clarifies that the jump operators induce transitions between the single-particle localized states.  The unitary evolution on the other hand dephases the initial state leading to a diagonal density matrix in the basis $\tilde{n}_\a$. 
Hence, the equation of motion for $\tilde{n}_\a$ generated by  Eq.~(\ref{eq:lindblad}) is a classical rate equation with transition rates set by the  above jump operators, 
\begin{equation}
  \dot{\tilde{n}}_{\alpha} = \gamma\sum_{\beta} (w_{\alpha\beta} \tilde{n}_{\beta} - w_{\beta\alpha} \tilde{n}_{\alpha}) 
  \label{eq:eom-ni-diag}
\end{equation}
with $w_{\alpha\beta} = \sum_i|\phi^{\phantom{*}}_{\alpha i} \phi_{\beta i}^*|^2=\sum_i|\langle \alpha|n_i|\beta\rangle|^2=w_{\b\a}$. 
The random hopping between sites relaxes density modulations that may have existed in the initial state.
\begin{figure}[t]
  \centering
  \includegraphics[width=0.45\textwidth]{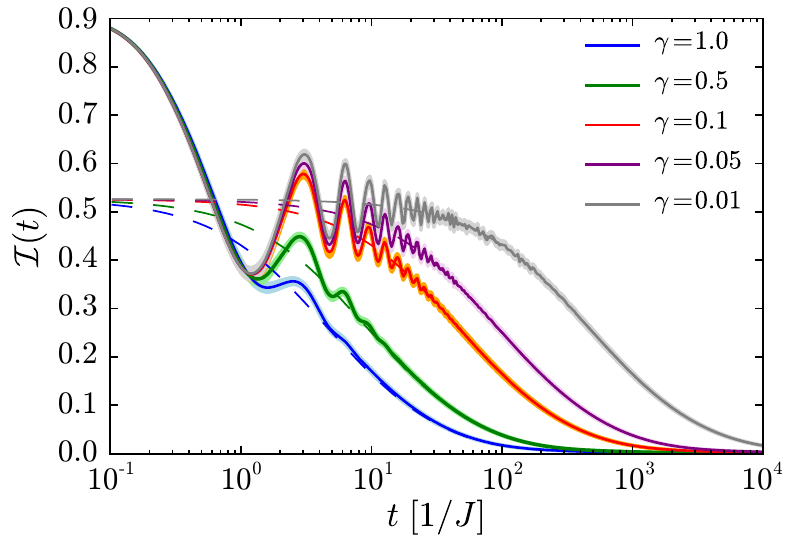}
  \caption{(Color online) Decay of an initially prepared imbalance $\mathcal{I}_0\sim 0.9$ for a disorder strength $h=10J$ under dephasing, i.e., $L_i=n_i$, for a system of size $N=20$ for $150$ disorder realizations. The shaded regions denote standard errors.}
  \label{fig:anderson-decay}
\end{figure}

Figure~\ref{fig:anderson-decay} shows a comparison of the result of the effective theory (dashed lines) with the exact solution available for the non-interacting problem (solid lines). Generally, Eq. \eqref{eq:lindblad} can be used to generate an infinite hierarchy of coupled equations for the $n$-point correlation functions. However, for quadratic Hamiltonians and the jump operators of interest here, the equations for the two-point functions $G_{\a\b}=\av{\tilde{c}\yd_\a\tilde{c}^{\pdag}_\b}$ are closed. For the dephasing bath they yield
\begin{equation}
  \dot{G}_{\alpha\beta} = i(\e_\a-\e_\b)G_{\alpha\beta}  +  \gamma \sum_{\mu\nu} w_{\alpha\mu}^{\beta\nu} G_{\mu\nu} -\gamma G_{\alpha\beta}  \label{eq:eom-ni}
\end{equation}
with the transition rates $w_{\alpha\mu}^{\beta\nu}= \sum_i \phi^{\phantom{*}}_{\alpha i}\phi_{\mu i}^* \phi^{\phantom{\star}}_{\nu i}\phi_{\beta i}^*$.
From Eq.~\eqref{eq:eom-ni}, we see that the off-diagonal terms are indeed averaged to zero because of the rapid oscillations with frequency of order $\e_\a-\e_\b\sim h$, the disorder strength.  The residual diagonal components then follow Eq. (\ref{eq:eom-ni-diag}).

\begin{figure}[t]
  \centering
  \includegraphics[width=0.45\textwidth]{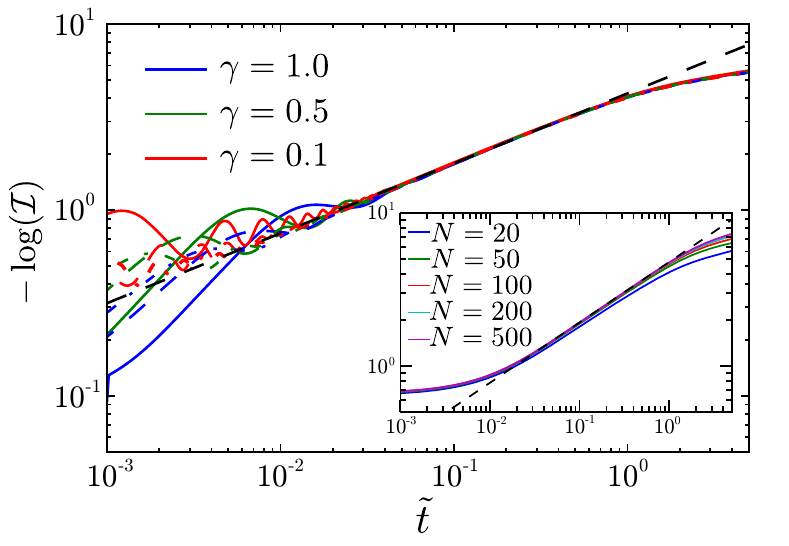}
  \caption{(Color online) Collapse of the imbalance decay for disorder strengths $h/J=10, 15, 20$ (solid, dashed, and dash-dotted lines, 150 disorder realizations each) under dephasing with $\tilde{t} = t \gamma (J/h)^2$. The dashed black line denotes $\exp[-(\tilde{t}/t_0)^{\beta}]$ with $\beta\approx 0.38$ and $t_0\approx 0.02$. The inset shows the size dependence using the diagonal rate equations.}
  \label{fig:anderson-scaling}
\end{figure}

In a strongly disordered system the transition rates are dominated by the spatial overlaps of nearest-neighbor states $w_{\a,\a+1}\sim (J/h)^2$. Hence, we expect the long-time behavior of $\mathcal{I}(t)$ taken for different parameter values $\gamma$ and $h$ to collapse to a single universal curve in the strong-disordered regime when plotted as a function of the rescaled time $\tilde{t}=t\gamma (J/h)^2$. This collapse is demonstrated in Fig.~\ref{fig:anderson-scaling}, showing the decay of the imbalance for different disorder strengths $h/J=10,15,20$. We have plotted $-\log\mathcal{I}(\tilde{t})$ in this figure in order to better identify the form of the decay function. The result appears like a stretched exponential decay $\mathcal{I} \sim \exp(-(t/t_0)^{\beta})$, with $\beta\approx 0.38$ and $t_0\approx0.02$, that persists for about two orders of magnitude in time. This is broadly consistent with the results reported in a recent numerical study \cite{levi:2015tmp}.

Such a stretched exponential decay is known from systems exhibiting a distribution of time scales and can be understood as a superposition of exponential decays~\cite{kohlrausch:1854}. In our system this could be a result of the imbalance decaying with different rates at distinct parts of the system because of fluctuations in the disorder strength. Assuming the imbalance has slow spatial variations throughout the decay the equation of motion for the local imbalance has a well-defined continuum limit (see SI) given by 
\begin{equation}
  \frac{d}{dt}\mathcal{I}(x, t) \approx -\Gamma(x) \mathcal{I}(x, t) +\text{gradients}.
  \label{eq:coarse-grained}
\end{equation}
Here, $\Gamma(x)$ is a local, effective decay rate. Deep in the localized phase, $\Gamma(x)$ is dominated by short-range hopping rates, $\Gamma\sim 4 \gamma [J/\Delta(x)]^2$, where $\Delta(x)$ is the energy difference between nearby sites at location $x$. After coarse graining we expect $\Delta$ to have a Gaussian distribution with variance $\sigma^2 = \sigma^2_0 h^2<2h^2/3$.  

First solving for the local imbalance $\mathcal{I}(x,t)=\mathcal{I}_0\exp[-\Gamma(x) t]$ with $\mathcal{I}_0$ the initial imbalance, we can obtain the global imbalance from the integral
\be
  \mathcal{I}(t)\approx\mathcal{I}_0 \int dx  e^{-\Gamma(x) t} 
  ={\mathcal{I}_0\over \sqrt{2\pi\sigma^2}}\int d\Delta \,\,e^{-{\D^2\over 2\sigma^2} - {4J^2\over \D^2}\gamma t}.
  \label{eq:integratedI}
\ee
The saddle-point approximation to this integral gives a stretched exponential decay $\mathcal{I}(t)=\mathcal{I}_0\exp(-\sqrt{\tilde{t}/t_0})$, where $\tilde{t}=t \gamma (J/h)^2$ is the scaled time defined above and $t_0=\sigma_0^2 / 8$. The numerical value of $t_0\approx 0.02$ thus corresponds to a variance for $\Delta$ of $\sigma^2\approx h^2/6$.
The small change in exponent, which was fit to 0.38 in Fig.~\ref{fig:anderson-scaling} for $N=20$, shows only weak dependence on system size (see inset Fig.~\ref{fig:anderson-scaling}) and may be due to longer-range hopping or corrections from the gradient terms. 

Next, we consider dissipation in the form of single-particle loss. Such a bath is implemented by the jump operators $L_i = c_i=\sum_\alpha\phi^*_{\a i}\tilde{c}^{\phantom{*}}_\a$. In this case, the classical effective model leads to a trivial rate equation
\be
\dot{\tilde{n}}_\a=-\g\tilde{n}_{\alpha}.
\ee
The exact equations of motion for the two-point functions are also simpler than in the dephasing bath,
\begin{equation}
  \dot{G}_{\alpha\beta} = i(\epsilon_{\alpha}-\epsilon_{\beta})G_{\alpha\beta} - \gamma G_{\alpha\beta}.
  \label{eq:eom-two-point}
\end{equation}
Again, on a timescale $\sim 1/h$ off-diagonal terms die off  and the effective model and exact dynamics coincide. 

Since the decay due to single-particle loss in the non-interacting system does not induce hopping between different localized orbitals and is symmetric with respect to even and odd sites, it does not lead to decay of the (normalized) imbalance.
Only the total particle number relaxes with a trivial exponential decay.

\emph{Interacting system --}
In the full MBL phase, where all eigenstates are localized, the Hamiltonian
\begin{equation}
  H = \sum_\alpha\tilde{\epsilon}_\alpha \tilde{n}_\alpha + \sum_{\alpha \beta}U_{\alpha \beta}\tilde{n}_\alpha \tilde{n}_\beta + \cdots
  \label{eq:Heff}
\end{equation}
only contains products of the quasi-local integrals of motion $\tilde{n}_{\alpha}$~\cite{serbyn:2013, huse:2014, ros:2015, chandran:2015}.
In order to write the Lindblad equation~\eqref{eq:lindblad} as a classical rate equation~\eqref{eq:classicalrate}, we have to express the fermion operators $c_i$ in terms of the LIOM operators $\tilde{c}_\alpha$, which for the interacting case become non-trivial. 

We first consider the weak-coupling limit, $U\ll \max(h,J)$. For a perturbative expansion of the LIOMs we express the Hamiltonian Eq.~\eqref{eq:HIntAnderson} using the operators $c_\alpha$ that diagonalize the problem without interactions
\begin{equation}
  H = \sum_{\alpha}\epsilon_\alpha n_\alpha + \sum_{\alpha, \beta} U_{\alpha\beta}n_\alpha n_\beta + \sum{}' U_{\alpha\beta\gamma\delta}c^\dag_\alpha c^\dag_\beta c^\pdag_\gamma c^\pdag_\delta,
  \label{eq:fullH}
\end{equation}
where $\sum{}'$ contains no diagonal terms $U_{\alpha\beta} = U_{\alpha\beta\beta\alpha}$ and 
$U_{\alpha\beta\gamma\delta} =U \sum_i \phi^{\phantom{*}}_{i \alpha}\phi^{\phantom{*}}_{i+1\beta}\phi_{i+1\gamma}^*\phi_{i \delta}^*.$
Now, by requiring  $[H, \tilde{n}_\alpha]=O(U^2/h^2)$, we obtain  
\begin{equation}
  \tilde{c}_\alpha \!\sim\! c_\alpha \!+ \sum_{\beta\gamma\delta}\!\frac{U_{\alpha\beta\gamma\delta} + U_{\beta\alpha\gamma\delta}}{\epsilon_\alpha \!+\! \epsilon_{\beta} \!-\! \epsilon_{\gamma} \!-\! \epsilon_{\delta}} \, c^\dag_\beta c^\pdag_\gamma c^\pdag_\delta=c_\alpha\! + \sum_{\beta\gamma\delta}\!A^\alpha_{\beta\gamma\delta} \, c^\dag_\beta c^\pdag_\gamma c^\pdag_\delta.
  \label{eq:tildec}
\end{equation}
For the validity of this expansion we must have no resonant terms in the above sum. This requirement is identical to the MBL criterion, formulated by \textcite{basko:2006}, of no incoherent decay of a particle from state $\alpha$.
 
Inverting Eq.~\eqref{eq:tildec} we find the equations of motion for the LIOMs using Eq.~\eqref{eq:classicalrate}. For the case of dephasing, already the non-interacting system exhibits diffusion and the interaction effects are only sub-leading. Specifically in the regime $h\gg J\gg U$ the dissipation induced hopping rates behave as $\Gamma\sim (J/h)^2\left(1+ A(U/h)^2\right)$, with $A$ a numerical constant of order 1 (see SI). 

\begin{figure}[t]
  \centering
  \includegraphics[width=0.45\textwidth]{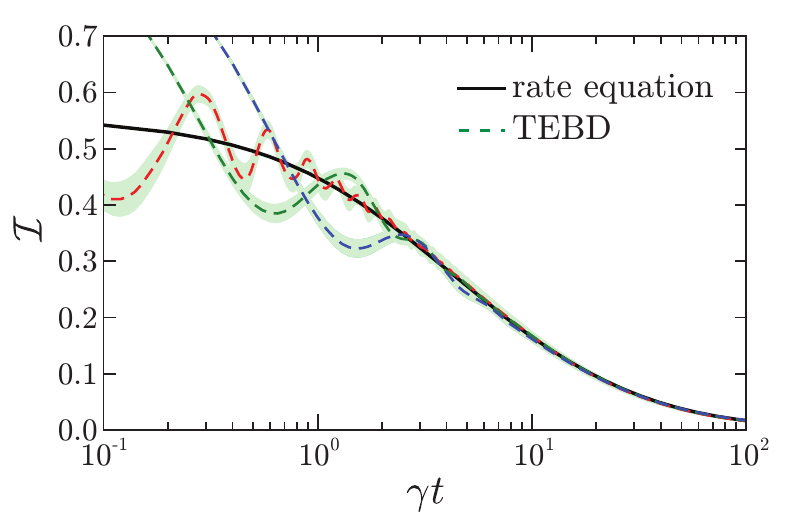}
  \caption{Comparison of the dynamics from the classical rate equation (solid line, 250 disorder realizations) and the numerical simulation (dashed lines, $\gamma/J=0.1, 0.5, 1.0$, 100 disorder realizations) for a system of size $N=20$, $U=0.5J$, and $h=10J$. The shaded regions denote standard errors.}
  \label{fig:interacting-dephasing}
\end{figure}

We solve the effective classical rate equations obtained for the dephasing process and compare  with a simulation of the quantum Lindblad equation ~\eqref{eq:lindblad} using the time-evolving block decimation (TEBD) scheme for matrix product operators~\cite{vidal:2004,zwolak:2004, orus:2008, schollwock:2011}.
Fig.~\ref{fig:interacting-dephasing} shows excellent agreement between the effective model and simulations of  a 20-site chain.

\begin{figure}[t]
  \centering
  \includegraphics{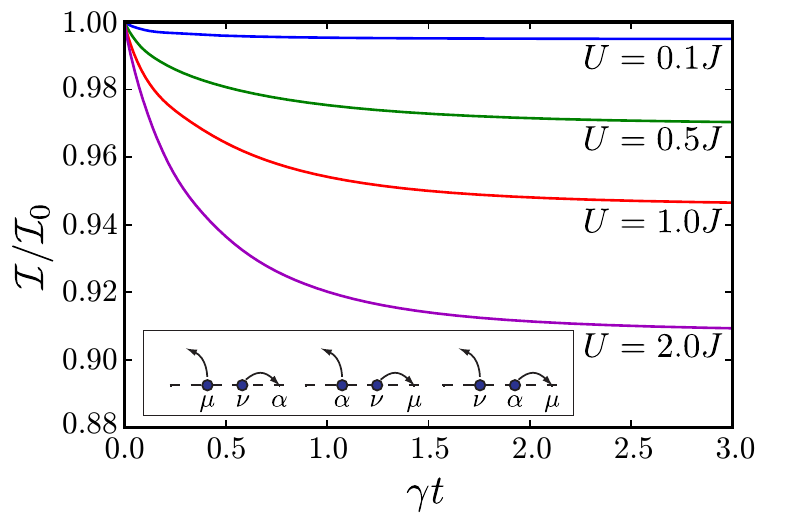}
  \caption{Imbalance decay in the presence of interactions for particle loss calculated using the classical rate equations for a system of size $N=20$, $h=10J$, $\gamma=0.1J$, and 250 disorder realizations. The inset shows the induced hopping processes that lead to diffusion as described by Eq.~\eqref{eq:liom_eom}.}
  \label{fig:ind-hopping}
\end{figure}

The particle loss in the interacting case is governed by the rate equation
\begin{multline}
  \dot{\tilde{n}}_\alpha \approx -\gamma \tilde{n}\alpha + \sum_{\mu\nu} W_{\alpha}^{\mu\nu} (1-\tilde{n}_\alpha)\tilde{n}_\mu\tilde{n}_\nu\\
  - 2 \sum_{\mu\nu} W_\mu^{\alpha\nu} (1-\tilde{n}_\mu)\tilde{n}_\alpha\tilde{n}_\nu
  \label{eq:liom_eom}
\end{multline}
with $W_\alpha^{\mu\nu}=\gamma \sum_\tau |A^\tau_{\alpha\mu\nu}|^2$.
In the interacting system, particle loss additionally induces particle hopping (see inset Fig.~\ref{fig:ind-hopping}), which in turn leads to diffusion. Thus, the dynamics for particle loss in the interacting case differs clearly from the non-interacting case. Only when the system becomes so dilute as to effectively be non-interacting, the imbalance stabilizes, as shown in Fig.~\ref{fig:ind-hopping}.

Finally, we consider the strong-coupling limit $U\gg h\gg J$. For simplicity, we study the dynamics of the initial state $|\Omega\rangle = \Pi_i c^\dag_{2i}|0\rangle$, the perfect density wave. 
In the limit $U\to \infty$ this initial state is a stationary state of the dynamics if we include only the dephasing bath, hence the decay rate of the imbalance must vanish in this limit. When $U$ is large but finite we can obtain the LIOMs using perturbation theory in $J/U$ with $U$ having a similar role as the disorder $h$ had at weak coupling. We obtain a decay rate that scales as $\gamma(J/U)^2$ to leading order in the small parameter. This should be compared with the weak coupling rate $\sim\gamma(J/h)^2 \left[1+A(U/h)^2\right]$. Thus, as we vary $U$ from weak to strong coupling the maximal decay rate is expected to occur around $U\sim h$~\cite{bera:2015, mascarenhas:2015tmp}. 

It is also interesting to consider the effect of single particle loss in the strong interaction limit. For simplicity we take $U\to \infty$ and consider the dynamics at sufficiently short times when there are still few holes in the density wave. Then, the initial state in the vicinity of the lost particle can be taken to be the perfect density wave. To lowest order in the hopping, the final state after the particle loss is $|\Psi_{2i}\rangle = \tilde{c}_{2i} |\Omega\rangle$ with the renormalized single-fermion operator
\be
\tilde{c}_{2i}=c_{2i}+W^i_{-}c^\dag_{2i-1}c_{2i-2}c_{2i} + W^i_{+}c^\dag_{2i+1}c_{2i+2}c_{2i}
\ee
and $W^i_{\pm} = J/(h_{2i\pm2}-h_{2i\pm1})$. The first-order terms result in a decay of the imbalance at an initial rate that scales as $\gamma (J/h)^2$ to be contrasted with the decay rate $\sim \gamma (UJ/h^2)^2$ obtained for weak interactions $U,J\ll h$. Hence, we find an enhancement of the decay of the imbalance by interactions that, for the case of single-particle loss, persists all the way to the limit of infinite $U$.

\emph{Conclusions  --}
We have presented a general framework to study the dynamics in MBL systems coupled weakly to a dissipative Markovian bath. Making use of the local integrals of motion characterizing the MBL phase, we have mapped the quantum evolution to much simpler classical rate equations. The form of these equations depends on the type of coupling to the bath and the intrinsic interactions in the Hamiltonian. Compared to the quantum evolution the effective classical equations are feasible to tackle numerically and, with further approximations, even analytically. 

Our effective model provides a simple explanation for the stretched exponential decay of observables in the localized phase. The dynamics of MBL systems weakly coupled to a bath is seen, from this perspective, to be similar to that of a classical glass \cite{binder:2011}. Further study is needed to understand how different classes of disorder affect the pattern of relaxation of observables (see SI for comparison with the case of a quasi-periodic potential studied in recent experiments~\cite{schreiber:2015}).

Our scheme of mapping the quantum dynamics to classical rate equations relies on the fact that the action of a local operator does not cause resonant transitions if the system is deep in the localized state. This follows from the large offset in energy associated with nearby integrals of motion. However, if the system is closer to the delocalization transition then resonances start to play an increasingly important role. It would therefore be very interesting to investigate how the relaxation behavior changes as the system is tuned toward the transition. This could give important information about the nature of the critical-thermal regions that destabilize the MBL state \cite{Vosk2015,Potter2015}.

\section{Acknowledgments}

Useful discussions with Immanuel Bloch, Pranjal Bordia, Andrew Daley, Markus Heyl, Henrik Lueschen, Evert van Nieuwenburg, Ulrich Schneider, and Lukas M.~Sieberer are gratefully acknowledged. We thank the ERC for financial support through the UQUAM synergy grant.  MHF acknowledges additional support from the Swiss Friends of the Weizmann Institute of Science.

\begin{widetext}
\section*{Supplementary Information}
\renewcommand{\thesection}{S\arabic{section}}    
\renewcommand{\thefigure}{S\arabic{figure}}
\renewcommand{\theequation}{S\arabic{equation}} 

\setcounter{figure}{0}
\setcounter{equation}{0}

\section{Origin of the stretched exponential decay}
In this section we provide a simple explanation for the observed stretched exponential relaxation of the imbalance.
As a first step we derive an equation of motion that describes the spatial variations in the relaxation of the imbalance.
As a starting point we take equation (4) of the main text for the local densities $\tilde{n}_\a$. Deep in the localized phase we can take $w_{\a\b}$ to connect only nearest neighbors, i.e. $\b=\a\pm 1$. 

We now replace the local density $\tilde{n}_\a$  that has a dominant period two modulation on the lattice by the variable $f(x_\a)$ defined through
\be
\tilde{n}_\a = n_0+ e^{i\pi x_\a} f(x_\a).
\ee
Physically, $f(x)$ is the local amplitude of the period-2 charge density wave. The local normalized imbalance is $\mathcal{I}(x)=f(x)/n_0$. 
Because $f(x)$ is slowly varying in space, it facilitates a continuum limit of the rate equation
\be
\dot{f}=-\Gamma(x)f-{1\over 4}\Gamma(x)\partial_x^2 f+{1\over 4}\partial_x\Gamma(x) \partial_x f,
\ee 
where $\Gamma(x_\a)\equiv 2(w_{\a,\a+1}+w_{\a-1,\a})$. 
Neglecting the gradients of the smooth function $f(x,t)$ we find 
$f(x,t)\approx n_0\mathcal{I}_0\exp[-\Gamma(x) t]$. The relaxation of the imbalance then follows from
\be
\mathcal{I}(t)={\mathcal{I}_0\over L}\int_0^L dx e^{-\Gamma(x) t}.
\ee

We have seen in the main text that in the strongly localized state the local relaxation rate $\Gamma(x)$ depends on the  energy offsets $\Delta$ between neighboring sites as $\Gamma(x)=4\gamma [J/\Delta(x)]^2$. After coarse graining we expect the energy offsets to follow a gaussian distribution
\be
P(\Delta)={1\over \sqrt{2\pi \sigma^2}}\,\,e^{-{\D^2\over 2 \sigma^2}},
\ee
where $\sigma^2 = \sigma_0^2 h^2$ is the variance of $\D$, the offset between two adjacent sites, and  $h$ is the on-site disorder strength. Note that due to the coarse graining, we expect $\sigma_0^2 < 2/3$, the value corresponding to the variance of the probability distribution of the local energy differences. 

Putting all this together we can convert the integral over $x$ to one over $\Delta$ to obtain
\be
\mathcal{I}(t)={\mathcal{I}_0\over \sqrt{2\pi \sigma_0^2 h^2}}\int d\Delta \,\,e^{-{\D^2\over 2\sigma_0^2h^2} - {4J^2\over \D^2}\gamma t}.
\ee
Finally, a saddle-point evaluation of this integral yields
\be
\mathcal{I}(t)\sim  e^{-\sqrt{ 8(J/ h)^2\gamma t/\sigma_0^2\,}}=e^{-\sqrt{\tilde{t}/t_0}},
\ee
where $\tilde{t}=t\gamma (J/ h)^2$ is the scaled time defined in the main text and $t_0 = \sigma_0^2 / 8$. For the numerical result of $t_0\approx0.02$, we would require $\sigma^2_0\approx 1/6$ which corresponds to coarse graining over order four lattice sites. 

\section{Comparison to quasi-periodic `disorder'}
To emphasize how the type of disorder influences the form of the decay of conserved quantities when coupled to a bath, we compare the non-interacting case of dephasing for the random disorder to the quasi-periodic disorder of the Aubry-Andr\'e model~\cite{aubry:1980}, 
\begin{equation}
	H = -J\sum_i (c^{\dag}_i c^{\phantom{\dag}}_{i+1} + h.c.) + \sum_i V_i n_i,
\end{equation}
where $V_i = h \cos(2\pi\beta i + \phi)$ with $\beta = (\sqrt{5}-1)/2$ and $\phi$ a random phase. As shown in Fig.~\ref{fig:comp}, this type of disorder not only localizes more strongly for $h=10$, but also changes the exponent $\beta$ and time scale $t_0$ to $\beta\approx0.47$ and $t_0\approx0.014$.

\begin{figure}[b]
  \centering
  \includegraphics{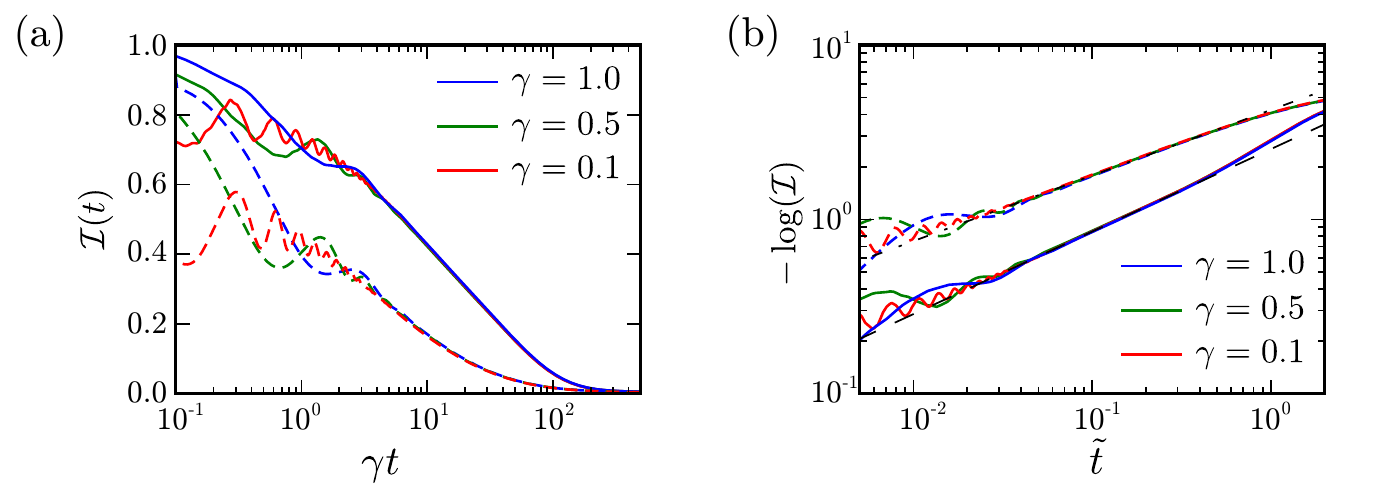}
  \caption{(Color online) Comparison of the decay of the imbalance for random disorder (dashed lines) and a quasi-periodic potential (solid lines) for $h=10J$ and $U=0$. (a) The decay of the charge density wave as a function of $\log(\gamma t)$, and (b) $- \log (\tilde{t})$ vs. $\tilde{t}$ in log-log scale to emphasize the different stretched exponential form.}
  \label{fig:comp}
\end{figure}

\section{Rate equation for dephasing bath}
To lowest order in $U$, we can trivially invert Eq.~(13) of the main text to express the jump operators in terms of the operators for the local integrals of motion. First, 
\begin{equation}
  c_\alpha \sim \tilde{c}_\alpha - \sum_{\beta\gamma\delta} A^\alpha_{\beta\gamma\delta} \tilde{c}^\dag_\beta \tilde{c}^\pdag_\gamma \tilde{c}^\pdag_\delta,
  \label{eq:cinverted}
\end{equation}
which we use to find the jump operators
\begin{equation}
  L_i = c^{\dag}_i c^\pdag_i  =\sum_{\alpha\beta}\phi^{\phantom{*}}_{\alpha i}\phi^{*}_{\beta i} c_\alpha^{\dag}c^{\pdag}_{\beta} \approx \sum_{\alpha\beta}\phi^{\phantom{*}}_{\alpha i}\phi^{*}_{\beta i} \tilde{c}^{\dag}_{\alpha}\tilde{c}^{\pdag}_{\beta} - \sum_{\alpha\beta}\sum_{\mu\nu\tau}\phi^{\phantom{*}}_{\alpha i}\phi^{*}_{\beta i} A^{\beta}_{\mu\nu\tau}\tilde{c}^{\dag}_{\alpha}\tilde{c}^{\dag}_{\mu}\tilde{c}^{\pdag}_{\nu}\tilde{c}_{\tau}^{\pdag} - \sum_{\alpha\beta}\sum_{\mu\nu\tau}\phi^{\phantom{*}}_{\alpha i}\phi^{*}_{\beta i} (A^{\alpha}_{\mu\nu\tau})^{*}\tilde{c}^{\dag}_{\nu}\tilde{c}^{\dag}_{\tau}\tilde{c}^{\pdag}_{\mu}\tilde{c}_{\beta}^{\pdag}.
  \label{eq:full-jump-intdeph}
\end{equation}
We therefore obtain
the classical rate equations
\begin{eqnarray}
  \dot{\tilde{n}}_\alpha &\approx&\gamma\sum_{\beta} [w_{\alpha\beta}\tilde{n}_{\beta}(1-\tilde{n}_{\alpha}) - w_{\beta\alpha}\tilde{n}_\alpha(1-\tilde{n}_{\beta})] \nonumber\\
 && + \gamma \sum_{\beta\mu}\left[(W_{\alpha\beta}^{\mu\mu}+W^{\alpha\beta}_{\mu\mu})\tilde{n}_{\mu}\tilde{n}_{\beta}(1-\tilde{n}_{\alpha})-(W_{\beta\alpha}^{\mu\mu}+W^{\beta\alpha}_{\mu\mu})\tilde{n}_{\mu}\tilde{n}_{\alpha}(1-\tilde{n}_{\beta})\right]\nonumber\\
 &&+ \gamma \sum_{\beta\mu\nu}\left[(W_{\alpha\beta}^{\mu\nu}+W_{\beta\alpha}^{\mu\nu})(1-\tilde{n}_{\alpha})(1-\tilde{n}_{\beta}) \tilde{n}_{\mu}\tilde{n}_{\nu} - 2W^{\alpha\beta}_{\mu\nu}(1-\tilde{n}_{\mu})(1-\tilde{n}_{\nu}) \tilde{n}_{\alpha}\tilde{n}_{\beta}\right].
  \label{eq:full-eom-intdeph}
\end{eqnarray}
with $w_{\alpha\beta} = \sum_i|\phi^{\phantom{*}}_{\alpha i}\phi^{*}_{\beta i}|^2$ and
\begin{equation}
  W_{\alpha\beta}^{\mu\nu} = \sum_i \left|\sum_{\tau}\phi^{\phantom{*}}_{\alpha i}\phi^{*}_{\tau i} A^{\tau}_{\mu\nu\beta}\right|^2.
  \label{eq:fullWabcd}
\end{equation}
Note that the first term of the right-hand side of Eq.~\eqref{eq:full-eom-intdeph} is identical to the rate equation~(13) for the non-interacting case, which already leads to diffusion and a loss of the imbalance. 
For $h\gg J$, the next-order contributions come from the second line which describes an `assisted hopping', leading to a correction  $\sim(JU/h^2)^2$, such that the dissipation induced hopping rates are of the form $\Gamma\sim (J/h)^2\left(1+ A(U/h)^2\right)$, with $A$ a numerical constant of order 1.
Indeed, the functional form stays essentially the same all the way up to $U/h\sim 1$, with the same parameters, as shown in Fig.~\ref{fig:betas}.
\begin{figure}[h]
  \centering
  \includegraphics{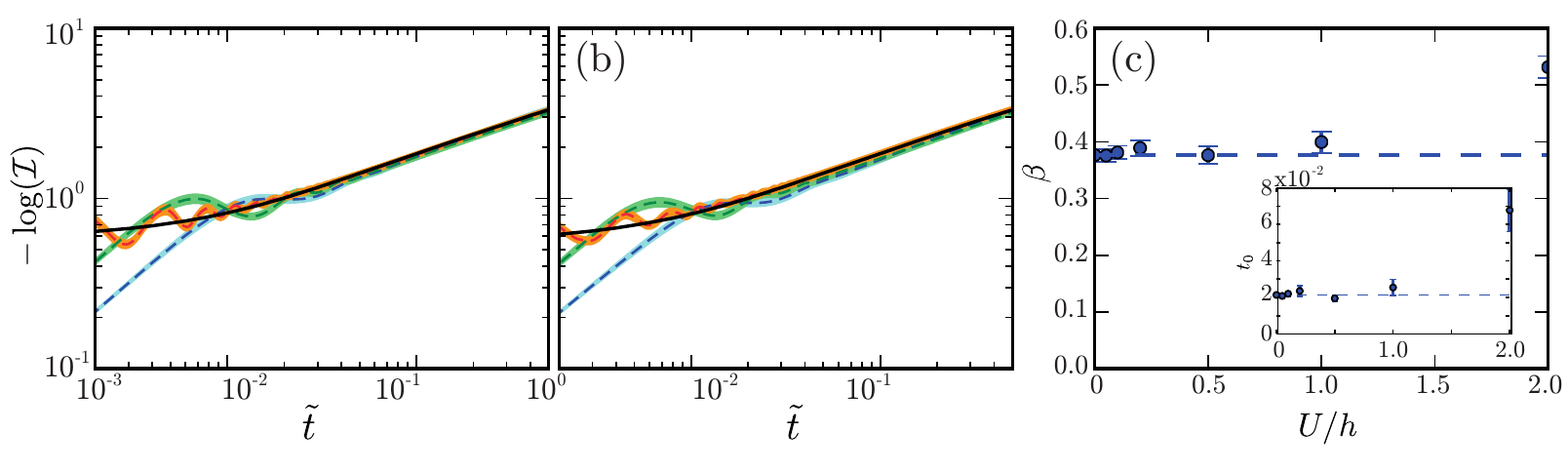}
  \caption{(Color online) Comparison of decay of the imbalance for dephasing bath for classical rate equation (solid line) and TEBD (dashed lines) for (a) $U=1.0J$ and (b) $U=2.0J$. (c) The stretched exponential $\beta$ for the decay of the imbalance $\mathcal{I} \sim \exp[-(t/t_0)^\beta]$ for $h=10$ and $50$ disorder realizations for various $U$. The inset shows the corresponding time scale $t_0$.}
  \label{fig:betas}
\end{figure}

\end{widetext}

\end{document}